\documentclass[12pt]{article}
\usepackage{amssymb,textcomp,amsmath,graphicx}

\begin{document}
\def\gsim{\:\raisebox{-0.5ex}{$\stackrel{\textstyle>}{\sim}$}\:}
\def\lsim{\:\raisebox{-0.5ex}{$\stackrel{\textstyle<}{\sim}$}\:}
\renewcommand{\thefootnote}{\fnsymbol{footnote}}

\thispagestyle{empty}
\begin{center}
\Large {\bf Mini--Review of Dark Matter: 2012}\footnote{This review
 summarizes the state of the field as of the end of November 2011;
 later developments are not covered. Note also that, as per PDG
 guidelines, we have tried to keep the list of references as short as
 possible; apologies to the many authors whose papers could not be cited.}

\vspace*{5mm}
\large{Manuel Drees} \\
\textit{Physikalisches Institut and Bethe Center for Theoretical Physics,\\ 
 Universit\"{a}t Bonn, Nussallee 12, 53115 Bonn, Germany}

\vspace*{3mm}
\large{Gilles Gerbier}\\
\textit{Centre d'Etudes de Saclay, F-91191 Gif-sur-Yvette Cedex,
  France}
\end{center}

\vspace*{1cm}

\begin{abstract}

  This is the mini--review on Dark Matter in the 2012 edition of the
  Particle Data Group's Review of Particle Properties. After briefly
  summarizing the arguments in favor of the existence of Dark Matter,
  we list possible candidates, ranging in mass from a fraction of an
  eV ({\it e.g.}, axions) to many solar masses ({\it e.g.}, primordial
  black holes), and discuss ways to detect them. The main emphasis is
  on Weakly Interacting Massive Particles (WIMPs). A large
  international effort is being made to detect them directly, or else
  to detect their annihilation products. We explain why we consider
  all claims to have established a positive signal for WIMPs in either
  direct or indirect detection to be premature. We also introduce the
  concept of a {\it WIMP safe} minimal mass; below this mass, the
  interpretation of a given direct search experiment depends strongly
  on the tail of the WIMP velocity distribution and/or on the
  experimental energy resolution.

\end{abstract}

\clearpage
\setcounter{page}{1}
\pagestyle{plain}

\section{Theory}
\label{darkmatAdg}

\subsection{Evidence for Dark Matter}

The existence of Dark ({\it i.e.}, non-luminous and non-absorbing)
Matter (DM) is by now well established \cite{dm_bertone,dm_hist}. The
earliest, and perhaps still most convincing, evidence for DM came from
the observation that various luminous objects (stars, gas clouds,
globular clusters, or entire galaxies) move faster than one would
expect if they only felt the gravitational attraction of other visible
objects. An important example is the measurement of galactic rotation
curves. The rotational velocity $v$ of an object on a stable Keplerian
orbit with radius $r$ around a galaxy scales like $v(r) \propto \sqrt{
  M(r) / r}$, where $M(r)$ is the mass inside the orbit. If $r$ lies
outside the visible part of the galaxy and mass tracks light, one
would expect $v(r) \propto 1/\sqrt{r}$. Instead, in most galaxies one
finds that $v$ becomes approximately constant out to the largest
values of $r$ where the rotation curve can be measured; in our own
galaxy, $v \simeq 240$ km$/$s at the location of our solar system,
with little change out to the largest observable radius. This implies
the existence of a {\it dark halo}, with mass density $\rho(r) \propto
1/r^2$, {\it i.e.}, $M(r) \propto r$; at some point $\rho$ will have to fall
off faster (in order to keep the total mass of the galaxy finite), but
we do not know at what radius this will happen. This leads to a lower
bound on the DM mass density, $\Omega_{\rm DM} \gsim 0.1$, where
$\Omega_X \equiv \rho_X / \rho_{\rm crit}$, $\rho_{\rm crit}$ being
the critical mass density ({\it i.e.}, $\Omega_{\rm tot} = 1$ corresponds to
a flat Universe).

The observation of clusters of galaxies tends to give somewhat larger values,
$\Omega_{\rm DM} \simeq 0.2$. These observations include measurements of the
peculiar velocities of galaxies in the cluster, which are a measure of their
potential energy if the cluster is virialized; measurements of the {\it X-ray}
temperature of hot gas in the cluster, which again correlates with the
gravitational potential felt by the gas; and---most directly---studies of
(weak) gravitational lensing of background galaxies on the cluster.

A particularly compelling example involves the bullet cluster
(1E0657-558) which recently (on cosmological time scales) passed
through another cluster.  As a result, the hot gas forming most of the
clusters' baryonic mass was shocked and decelerated, whereas the
galaxies in the clusters proceeded on ballistic
trajectories. Gravitational lensing shows that most of the total mass
also moved ballistically, indicating that DM self-interactions are
indeed weak \cite{dm_bertone}.

The currently most accurate, if somewhat indirect, determination of
$\Omega_{\rm DM}$ comes from global fits of cosmological parameters to
a variety of observations; see the Section on Cosmological Parameters
for details. For example, using measurements of the anisotropy of the
cosmic microwave background (CMB) and of the spatial distribution of
galaxies, ref. \cite{dm_wmap} finds a density of cold, non-baryonic matter
\begin{equation} \label{edm1} 
\Omega_{\rm nbm} h^2 = 0.112 \pm 0.006\ ,
\end{equation}
where $h$ is the Hubble constant in units of 100 km/(s$\cdot$Mpc).
Some part of the baryonic matter density \cite{dm_wmap},
\begin{equation} \label{edm2}
\Omega_{\rm b} h^2 = 0.022 \pm 0.001\ ,
\end{equation}
may well contribute to (baryonic) DM, {\it e.g.}, MACHOs \cite{dm_macho} or cold
molecular gas clouds \cite{dm_gas}. 

The DM density in the ``neighborhood'' of our solar system is also of
considerable interest. This was first estimated as early as 1922 by
J.H. Jeans, who analyzed the motion of nearby stars transverse to the
galactic plane \cite{dm_hist}. He concluded that in our galactic
neighborhood, the average density of DM must be roughly equal to that
of luminous matter (stars, gas, dust). Remarkably enough, the most
recent estimate, based on a detailed model of our galaxy constrained
by a host of observables including the galactic rotation curve, finds
a quite similar result for the smooth component of the local Dark
Matter density \cite{dm_locden}:
\begin{equation} \label{edm2a}
\rho_{\rm DM}^{\rm local} = (0.39 \pm 0.03) \frac {\rm GeV} {{\rm cm}^3}\ .
\end{equation}
This value may have to be increased by a factor of $1.2 \pm 0.2$ since
the baryons in the galactic disk, in which the solar system is
located, also increase the local DM density \cite{dm_locdenfac}. Small
substructures (minihaloes, streams) are not likely to change the local
DM density significantly \cite{dm_bertone}.

\subsection{Candidates for Dark Matter}

Analyses of structure formation in the Universe indicate that most DM
should be ``cold'' or ``cool'', {\it i.e.}, should have been
non-relativistic at the onset of galaxy formation (when there was a
galactic mass inside the causal horizon) \cite{dm_bertone}. This
agrees well with the upper bound \cite{dm_wmap} on the contribution of
light neutrinos to Eq.(\ref{edm1}),
\begin{equation} \label{edm3}
\Omega_\nu h^2 \leq 0.0062 \ \ \ 95\% \ {\rm CL}\ .
\end{equation}
Candidates for non-baryonic DM in Eq.(\ref{edm1}) must satisfy several
conditions: they must be stable on cosmological time scales (otherwise
they would have decayed by now), they must interact very weakly with
electromagnetic radiation (otherwise they wouldn't qualify as {\it
dark} matter), and they must have the right relic density. Candidates
include primordial black holes, axions, sterile neutrinos, and weakly
interacting massive particles (WIMPs).

Primordial black holes must have formed before the era of Big-Bang
nucleosynthesis, since otherwise they would have been counted in
Eq.(\ref{edm2}) rather than Eq.(\ref{edm1}). Such an early creation of
a large number of black holes is possible only in certain somewhat
contrived cosmological models \cite{dm_bh}.

The existence of axions \cite{dm_axrev} was first postulated to solve
the strong {\it CP} problem of QCD; they also occur naturally in
superstring theories.  They are pseudo Nambu-Goldstone bosons
associated with the (mostly) spontaneous breaking of a new global
``Peccei-Quinn'' (PQ) U(1) symmetry at scale $f_a$; see the Section on
Axions in this {\it Review} for further details. Although very light,
axions would constitute cold DM, since they were produced
non-thermally. At temperatures well above the QCD phase transition,
the axion is massless, and the axion field can take any value,
parameterized by the ``misalignment angle'' $\theta_i$. At $T \lsim 1$
GeV, the axion develops a mass $m_a$ due to instanton effects. Unless
the axion field happens to find itself at the minimum of its potential
($\theta_i = 0$), it will begin to oscillate once $m_a$ becomes
comparable to the Hubble parameter $H$. These coherent oscillations
transform the energy originally stored in the axion field into
physical axion quanta. The contribution of this mechanism to the
present axion relic density is \cite{dm_bertone} 
\begin{equation} \label{edm4} 
\Omega_a h^2 = \kappa_a \left( f_a / 10^{12} \ {\rm GeV}
\right)^{1.175} \theta_i^2 \ , 
\end{equation}
where the numerical factor $\kappa_a$ lies roughly between $0.5$ and a
few. If $\theta_i \sim {\cal O}(1)$, Eq.(\ref{edm4}) will saturate
Eq.(\ref{edm1}) for $f_a \sim 10^{11}$ GeV, comfortably above
laboratory and astrophysical constraints \cite{dm_axrev}; this would
correspond to an axion mass around 0.1 meV. However, if the
post-inflationary reheat temperature $T_R > f_a$, cosmic strings will
form during the PQ phase transition at $T \simeq f_a$. Their decay
will give an additional contribution to $\Omega_a$, which is often
bigger than that in Eq.(\ref{edm4}) \cite{dm_bertone}, leading to a smaller
preferred value of $f_a$, {\it i.e.}, larger $m_a$. On the other hand,
values of $f_a$ near the Planck scale become possible if $\theta_i$ is
for some reason very small.

``Sterile'' $SU(2) \times U(1)_Y$ singlet neutrinos with keV
masses \cite{dm_sterile} could alleviate the ``cusp/core
problem'' \cite{dm_bertone} of cold DM models. If they were produced
non-thermally through mixing with standard neutrinos, they would eventually
decay into a standard neutrino and a photon.

Weakly interacting massive particles (WIMPs) $\chi$ are particles with mass
roughly between 10 GeV and a few TeV, and with cross sections of approximately
weak strength. Within standard cosmology, their present relic density can be
calculated reliably if the WIMPs were in thermal and chemical equilibrium with
the hot ``soup'' of Standard Model (SM) particles after inflation. In this
case, their density would become exponentially (Boltzmann) suppressed at $T <
m_\chi$. The WIMPs therefore drop out of thermal equilibrium (``freeze out'')
once the rate of reactions that change SM particles into WIMPs or vice versa,
which is proportional to the product of the WIMP number density and the WIMP
pair annihilation cross section into SM particles $\sigma_A$ times velocity,
becomes smaller than the Hubble expansion rate of the Universe. After freeze
out, the co-moving WIMP density remains essentially constant; if the Universe
evolved adiabatically after WIMP decoupling, this implies a constant WIMP
number to entropy density ratio. Their present relic density is then
approximately given by (ignoring logarithmic corrections) \cite{dm_kt}
\begin{equation} \label{edm5}
\Omega_\chi h^2 \simeq const. \cdot \frac {T_0^3} {M_{\rm Pl}^3
\langle \sigma_A v \rangle} \simeq \frac { 0.1 \ {\rm pb} \cdot c}
{\langle \sigma_A v \rangle }\ .  
\end{equation}
Here $T_0$ is the current CMB temperature, $M_{\rm Pl}$ is the Planck
mass, $c$ is the speed of light, $\sigma_A$ is the total annihilation
cross section of a pair of WIMPs into SM particles, $v$ is the
relative velocity between the two WIMPs in their cms system, and
$\langle \dots \rangle$ denotes thermal averaging. Freeze out happens
at temperature $T_F \simeq m_\chi/20$ almost independently of the
properties of the WIMP. This means that WIMPs are already
non-relativistic when they decouple from the thermal plasma; it also
implies that Eq.(\ref{edm5}) is applicable if $T_R > T_F$. Notice that
the 0.1 pb in Eq.(\ref{edm5}) contains factors of $T_0$ and $M_{\rm
  Pl}$; it is, therefore, quite intriguing that it ``happens'' to come
out near the typical size of weak interaction cross sections.

The seemingly most obvious WIMP candidate is a heavy neutrino. However, an
SU(2) doublet neutrino will have too small a relic density if its mass exceeds
$M_Z/2$, as required by LEP data. One can suppress the annihilation cross
section, and hence increase the relic density, by postulating mixing between a
heavy SU(2) doublet and some sterile neutrino. However, one also has to
require the neutrino to be stable; it is not obvious why a massive neutrino
should not be allowed to decay.

The currently best motivated WIMP candidate is, therefore, the
lightest superparticle (LSP) in supersymmetric models \cite{dm_susy}
with exact R-parity (which guarantees the stability of the
LSP). Searches for exotic isotopes \cite{dm_exo} imply that a stable
LSP has to be neutral. This leaves basically two candidates among the
superpartners of ordinary particles, a sneutrino, and a
neutralino. The negative outcome of various WIMP searches (see below)
rules out ``ordinary'' sneutrinos as primary component of the DM halo
of our galaxy. (In models with gauge-mediated SUSY breaking, the
lightest ``messenger sneutrino'' could make a good WIMP
\cite{dm_messn}.) The most widely studied WIMP is therefore the
lightest neutralino. Detailed calculations \cite{dm_bertone} show that
the lightest neutralino will have the desired thermal relic density
Eq.(\ref{edm1}) in at least four distinct regions of parameter
space. $\chi$ could be (mostly) a bino or photino (the superpartner of
the U(1)$_Y$ gauge boson and photon, respectively), if both $\chi$ and
some sleptons have mass below $\sim 150$ GeV, or if $m_\chi$ is close
to the mass of some sfermion (so that its relic density is reduced
through co-annihilation with this sfermion), or if $2 m_\chi$ is close
to the mass of the {\it CP}-odd Higgs boson present in supersymmetric
models. Finally, Eq.(\ref{edm1}) can also be satisfied if $\chi$ has a
large higgsino or wino component.

Many non-supersymmetric extensions of the Standard Model also contain viable
WIMP candidates \cite{dm_bertone}. Examples are the lightest $T-$odd particle
in ``Little Higgs'' models with conserved $T-$parity, or ``techni-baryons''
in scenarios with an additional, strongly interacting (``technicolor'' or
similar) gauge group.

There also exist models where the DM particles, while interacting only
weakly with ordinary matter, have quite strong interactions within an
extended ``dark sector'' of the theory. These were motivated by
measurements by the PAMELA, ATIC and Fermi satellites indicating
excesses in the cosmic $e^+$ and/or $e^-$ fluxes at high energies.
However, these excesses are relative to background estimates that are
clearly too simplistic ({\it e.g.}, neglecting primary sources of
electrons and positrons, and modeling the galaxy as a homogeneous
cylinder). Moreover, the excesses, if real, are far too large to be
due to usual WIMPs, but can be explained by astrophysical sources. It
therefore seems unlikely that they are due to Dark
Matter \cite{dm_cirelli}. Similarly, claims of positive signals for
direct WIMP detection by the DAMA and, more recently, CoGeNT and
CRESST collaborations (see below) led to the development of
tailor-made models to alleviate tensions with null experiments. Since
we are not convinced that these data indeed signal WIMP detection, and
these models (some of which were quickly excluded by improved
measurements) lack independent motivation, we will not discuss them
any further in this Review.

Although thermally produced WIMPs are attractive DM candidates because
their relic density naturally has at least the right order of
magnitude, non-thermal production mechanisms have also been suggested,
{\it e.g.}, LSP production from the decay of some moduli fields
\cite{dm_moddec}, from the decay of the inflaton \cite{dm_infdec}, or
from the decay of ``$Q-$balls'' (non-topological solitons) formed in
the wake of Affleck-Dine baryogenesis \cite{dm_qball}.  Although LSPs
from these sources are typically highly relativistic when produced,
they quickly achieve kinetic (but not chemical) equilibrium if $T_R$
exceeds a few MeV \cite{dm_kineq} (but stays below $m_\chi/20$). They
therefore also contribute to cold DM. Finally, if the WIMPs aren't
their own antiparticles, an asymmetry between WIMPs and antiWIMPs
might have been created in the early Universe, possibly by the same
(unknown) mechanism that created the baryon antibaryon asymmetry. In
such ``asymmetric DM'' models \cite{dm_adm} the WIMP antiWIMP
annihilation cross section $\langle \sigma_A v \rangle$ should be
significantly larger than $1 \, {\rm pb} \cdot c$, {\it c.f.}
Eq.(\ref{edm5}).

Primary black holes (as MACHOs), axions, sterile neutrinos, and WIMPs
are all (in principle) detectable with present or near-future
technology (see below). There are also particle physics DM candidates
which currently seem almost impossible to detect, unless they decay;
the present lower limit on their lifetime is of order $10^{25}$ to
$10^{26}$ s for 100 GeV particles. These include the gravitino (the
spin-3/2 superpartner of the graviton) \cite{dm_bertone}, states from the
``hidden sector'' thought responsible for supersymmetry
breaking \cite{dm_messn}, and the axino (the spin-1/2 superpartner of
the axion) \cite{dm_bertone}.

\section{Experimental detection of Dark Matter}

\subsection{The case of baryonic matter in our galaxy}

The search for hidden galactic baryonic matter in the form of MAssive
Compact Halo Objects (MACHOs) has been initiated following the
suggestion that they may represent a large part of the galactic DM and
could be detected through the microlensing effect \cite{dm_macho}. The
MACHO, EROS, and OGLE collaborations have performed a program of
observation of such objects by monitoring the luminosity of millions
of stars in the Large and Small Magellanic Clouds for several
years. EROS concluded that MACHOs cannot contribute more than 8\%
to the mass of the galactic halo \cite{dm_nomacho}, while MACHO
observed a signal at 0.4 solar mass and put an upper limit of 40\%. 
Overall, this strengthens the need for non-baryonic DM, also
supported by the arguments developed above.

\subsection{Axion searches}
\label{darkmatCdg}

Axions can be detected by looking for $a \rightarrow \gamma$
conversion in a strong magnetic field \cite{dm_bertone}. Such a
conversion proceeds through the loop-induced $a \gamma \gamma$
coupling, whose strength $g_{a \gamma \gamma}$ is an important
parameter of axion models. There currently are two experiments
searching for axionic DM. They both employ high quality cavities. The
cavity ``Q factor'' enhances the conversion rate on resonance, {\it i.e.},
for $m_a (c^2 +v_a^2/2) = \hbar \omega_{\rm res}$. One then needs to
scan the resonance frequency in order to cover a significant range in
$m_a$ or, equivalently, $f_a$. The bigger of the two experiments, the
ADMX experiment \cite{dm_llnlax}, originally situated at the LLNL in
California but now running at the University of Washington, started
taking data in the first half of 1996. It now uses SQUIDs as
first-stage amplifiers; their extremely low noise temperature (1.2 K)
enhances the conversion signal. Published results \cite{dm_llnlres},
combining data taken with conventional amplifiers and SQUIDs, exclude
axions with mass between 1.9 and 3.53 $\mu$eV, corresponding to $f_a
\simeq 4 \cdot 10^{13}$ GeV, for an assumed local DM density of 0.45
GeV$/$cm$^3$, if $g_{a \gamma \gamma}$ is near the upper end of the
theoretically expected range. An about five times better limit on
$g_{a\gamma\gamma}$ was achieved \cite{dm_admxhi} for $1.98 \ \mu{\rm
  eV} \leq m_a \leq 2.18 \ \mu$eV, if a large fraction of the local DM
density is due to a single flow of axions with very low velocity
dispersion. The ADMX experiment is being upgraded by reducing the
cavity and SQUID temperature from the current 1.2 K to about 0.1
K. This should increase the frequency scanning speed for given
sensitivity by more than two orders of magnitude, or increase the
sensitivity for fixed observation time.

The smaller ``CARRACK'' experiment now being developed in Kyoto,
Japan \cite{dm_kyotax} uses Rydberg atoms (atoms excited to a very high state,
$n =111$) to detect the microwave photons that would result from axion
conversion. This allows almost noise-free detection of single photons. Their
ultimate goal is to probe the range between 2 and 50 $\mu$eV with sensitivity
to all plausible axion models, if axions form most of DM.

\subsection{Searches for keV Neutrinos}

Relic keV neutrinos $\nu_s$ can only be detected if they mix with the
ordinary neutrinos. This mixing leads to radiative $\nu_s \rightarrow
\nu \gamma$ decays, with lifetime $\tau_{\nu_s} \simeq 1.8 \cdot
10^{21} \ {\rm s} \cdot (\sin\theta)^{-2} \cdot \left( 1 \ {\rm keV}/
m_{\nu_s} \right)^5$, where $\theta$ is the mixing
angle \cite{dm_sterile}. This gives rise to a flux of mono-energetic photons
with $E_\gamma = m_{\nu_s} / 2$, which might be observable by {\it X-ray}
satellites. In the simplest case the relic $\nu_s$ are produced only by
oscillations of standard neutrinos. Assuming that all lepton-antilepton
asymmetries are well below $10^{-3}$, the $\nu_s$ relic density can then be
computed uniquely in terms of the mixing angle $\theta$ and the mass
$m_{\nu_s}$. The combination of lower bounds on $m_{\nu_s}$ from analyses of
structure formation (in particular, the Ly$\alpha$ ``forest'') and upper
bounds on {\it X-ray} fluxes from various (clusters of) galaxies exclude
this scenario if $\nu_s$ forms all of DM. This conclusion can be evaded if
$\nu_s$ forms only part of DM, and/or if there is a lepton asymmetry $\geq
10^{-3}$ (i.e. some 7 orders of magnitude above the observed
baryon-antibaryon asymmetry), and/or if there is an additional source of
$\nu_s$ production in the early Universe, e.g. from the decay of heavier
particles \cite{dm_sterile}.

\subsection{Basics of direct WIMP search}
\label{darkmatDdg}

As stated above, WIMPs should be gravitationally trapped inside
galaxies and should have the adequate density profile to account for
the observed rotational curves. These two constraints determine the
main features of experimental detection of WIMPs, which have been
detailed in the reviews in \cite{dm_bertone}.

Their mean velocity inside our galaxy relative to its center is
expected to be similar to that of stars, {\it i.e.}, a few hundred
kilometers per second at the location of our solar system. For these
velocities, WIMPs interact with ordinary matter through elastic
scattering on nuclei. With expected WIMP masses in the range 10 GeV to
10 TeV, typical nuclear recoil energies are of order of 1 to 100 keV.

The shape of the nuclear recoil spectrum results from a convolution of
the WIMP velocity distribution, usually taken as a Maxwellian
distribution in the galactic rest frame, shifted into the Earth rest
frame, with the angular scattering distribution, which is isotropic to
first approximation but forward-peaked for high nuclear mass
(typically higher than Ge mass) due to the nuclear form
factor. Overall, this results in a roughly exponential spectrum. The
higher the WIMP mass, the higher the mean value of the
exponential. This points to the need for low nuclear recoil energy
threshold detectors.

On the other hand, expected interaction rates depend on the product of
the local WIMP flux and the interaction cross section. The first term
is fixed by the local density of dark matter, taken as 0.39 GeV/cm$^3$
[see Eq.(\ref{edm2a})], the mean WIMP velocity, typically 220 km/s, the
galactic escape velocity, typically 544 km/s \cite{dm_escape} and the
mass of the WIMP. The expected interaction rate then mainly depends on
two unknowns, the mass and cross section of the WIMP (with some
uncertainty \cite{dm_locden} due to the halo model).  This is why the
experimental observable, which is basically the scattering rate as a
function of energy, is usually expressed as a contour in the WIMP
mass--cross section plane.

The cross section depends on the nature of the couplings.  For
non-relativistic WIMPs, one in general has to distinguish spin-independent and
spin-dependent couplings. The former can involve scalar and vector WIMP and
nucleon currents (vector currents are absent for Majorana WIMPs, {\it e.g.}, the
neutralino), while the latter involve axial vector currents (and obviously
only exist if $\chi$ carries spin).  Due to coherence effects, the
spin-independent cross section scales approximately as the square of the mass
of the nucleus, so higher mass nuclei, from Ge to Xe, are preferred for this
search. For spin-dependent coupling, the cross section depends on the nuclear
spin factor; used target nuclei include 
$\rm ^{19}$F, $\rm ^{23}$Na, $\rm ^{73}$Ge, $\rm ^{127}$I, 
$\rm ^{129}$Xe, $\rm ^{131}$Xe, and $\rm ^{133}$Cs.

Cross sections calculated in MSSM models \cite{dm_mssm} induce rates of at most
1~evt~day$^{-1}$~kg$^{-1}$ of detector, much lower than the usual radioactive
backgrounds. This indicates the need for underground laboratories to protect
against cosmic ray induced backgrounds, and for the selection of extremely
radio-pure materials.

The typical shape of exclusion contours can be anticipated from this
discussion: at low WIMP mass, the sensitivity drops because of the
detector energy threshold, whereas at high masses, the sensitivity
also decreases because, for a fixed mass density, the WIMP flux
decreases $\propto 1/m_\chi$. The sensitivity is best for WIMP masses
near the mass of the recoiling nucleus.

Two important points are to be kept in mind when comparing exclusion
curves from various experiments between them or with positive indications of
a signal.

For an experiment with a fixed nuclear recoil energy threshold, the lower is
the considered WIMP mass, the lower is the fraction of the spectrum to which
the experiment is sensitive. This fraction may be extremely small in some
cases. For instance CoGeNT \cite{dm_cogent2}, using a Germanium detector with
an energy threshold of around 2 keV, is sensitive to about 10 \% of the
total recoil spectrum of a 7 GeV WIMP, while for XENON100 \cite{dm_xenon100},
using a liquid Xenon detector with a threshold of 8.4 keV, this fraction is
only 0.05~\% (that is the extreme tail of the distribution), for the same WIMP
mass. The two experiments are then sensitive to very different parts of the
WIMP velocity distribution.

A second important point to consider is the energy resolution of the
detector. Again at low WIMP mass, the expected roughly exponential spectrum is
very steep and when the characteristic energy of the exponential becomes of
the same order as the energy resolution, the energy smearing becomes
important. In particular, a significant fraction of the expected spectrum
below effective threshold is smeared above threshold, increasing artificially
the sensitivity. For instance, a Xenon detector with a threshold of 8 keV and
infinitely good resolution is actually insensitive to a 7 GeV mass WIMP,
because the expected energy distribution has a cut-off at roughly 5 keV. When
folding in the experimental resolution of XENON100 (corresponding to a
photo statistics of 0.5 photoelectron per keV), then around 1 \% of the signal
is smeared above 5 keV and 0.05~\% above 8 keV. Setting reliable cross section
limits in this mass range thus requires a complete understanding of the
response of the detector at energies well below the nominal threshold.

In order to homogenize the reliability of the presented exclusion curves, and
save the reader the trouble of performing tedious (though easy to do)
calculations, we propose to set cross section limits only for WIMP mass above
a {\it ``WIMP safe"} minimal mass value defined as the maximum of 1) the mass
where the increase of sensitivity from infinite resolution to actual
experimental resolution is not more than a factor two, and 2) the mass where
the experiment is sensitive to at least 1 \% of the total WIMP signal recoil
spectrum. These recommendations are irrespective of the content of the
experimental data obtained by the experiments.

\subsection{Status and prospects of direct WIMP searches}

Given the intense activity of the field, readers interested in more details
than the ones given below may refer to \cite{dm_bertone}, as well as to
presentations at recent conferences \cite{dm_portal}.

The first searches have been performed with ultra-pure semiconductors
installed in pure lead and copper shields in underground
environments. Combining a priori excellent energy resolutions and very pure
detector material, they produced the first limits on WIMP searches
(Heidelberg-Moscow, IGEX, COSME-II, HDMS) \cite{dm_bertone}. Planned
experiments using several tens of kg to a ton of Germanium run at liquid
nitrogen temperature (designed for double-beta decay search)---GERDA,
MAJORANA---are based in addition on passive reduction of the external and
internal electromagnetic and neutron background by using segmented detectors,
minimal detector housing, close electronics, pulse shape discrimination and
large liquid nitrogen or argon shields. Their sensitivity to WIMP interactions
will depend on their ability to lower the energy threshold sufficiently, while
keeping the background rate small.

The use of so called Point Contact Germanium detectors, with a very
small capacitance allowing to reach sub-keV thresholds, has given rise
to new results. The CoGeNT collaboration \cite{dm_cogent_web} has
operated a single 440 g Germanium detector with an effective threshold
of 400 eV in the Soudan Underground Laboratory for 56
days \cite{dm_cogent2}. After applying a rise time cut on the pulse
shapes in order to remove the surface interactions known to suffer
from incomplete charge collection, the resulting spectrum below 4 keV
is said by the authors to exhibit an irreducible excess of events,
with energy spectrum roughly exponential, compatible with a light mass
WIMP in the 7-11 GeV range, and cross section around $10^{-4}$
pb. However, this conclusion crucially depends on the energy dependent
rise time cut applied to the data and a sizeable leaking of surface
events into the kept spectrum cannot be excluded.  The authors
acknowledge themselves that a possible instrumental effect, leading to
such an excess, is worth investigating. Nevertheless, considerable
attention has been paid to the WIMP interpretation, largely due to the
temptation to consider it as a confirmation of the low mass WIMP
DAMA/LIBRA solution, without channeling (see below). A recent
unpublished analysis, presented at the TAUP 2011 conference, indicates
a reduction of the claimed signal by a factor 10.  Further
results \cite{dm_cogent3} based on data accumulated during one year led
to the claim of a 2.8 sigma modulation said to be compatible with a
WIMP.  Here again, the claim is considerably weakened by the fact that
the amplitude of the curve describing the expected WIMP modulation in
the 0.5-3 keV bin is too high by roughly a factor 2 (or more, if the
unmodulated ``signal'' has to be reduced) and wrongly leads to the
conclusion that the modulation is compatible with a standard WIMP in a
standard halo. This is also noted in \cite{dm_cogent4}.

A new consortium, CDEX/TEXONO, plans to build a 10 kg array of small and very
low (200 eV) threshold detectors, and to operate them in the new Chinese
Jinping underground laboratory, the deepest in the world.

In order to make further progress in the reliability of any claimed
signal, active background rejection and signal identification
questions have to be addressed. This is the focus of a growing number
of investigations and improvements. Active background rejection in
detectors relies on the relatively small ionization in nuclear recoils
due to their low velocity. This induces a reduction (``quenching'') of
the ionization/scintillation signal for nuclear recoil signal events
relative to $e$ or $\gamma$ induced backgrounds. Energies calibrated
with gamma sources are then called ``electron equivalent energies''
(keVee unit used below). This effect has been both calculated and
measured \cite{dm_bertone}. It is exploited in cryogenic detectors
described later. In scintillation detectors, it induces in addition a
difference in decay times of pulses induced by $e/\gamma$ events vs
nuclear recoils. In most cases, due to the limited resolution and
discrimination power of this technique at low energies, this effect
allows only a statistical background rejection. It has been used in
NaI(Tl) (DAMA, LIBRA, NAIAD, Saclay NaI), in CsI(Tl) (KIMS), and Xe
(ZEPLIN-I) \cite{dm_bertone,dm_portal}. Pulse shape discrimination is
particularly efficient in liquid argon. Using a high energy threshold,
it has been used for an event by event discrimination by the WARP
experiment, but the high threshold also leads to a moderate signal
sensitivity. No observation of nuclear recoils has been reported by
these experiments.

Two experimental signatures are predicted for true WIMP signals. One is a
strong daily forward/backward asymmetry of the nuclear recoil direction, due
to the alternate sweeping of the WIMP cloud by the rotating Earth. Detection
of this effect requires gaseous detectors or anisotropic response
scintillators (stilbene). The second is a few percent annual modulation of the
recoil rate due to the Earth speed adding to or subtracting from the speed of
the Sun. This tiny effect can only be detected with large masses; nuclear
recoil identification should also be performed, as the otherwise much larger
background may also be subject to seasonal modulation.

The DAMA collaboration has reported results from a total of 6 years exposure
with the LIBRA phase involving 250 kg of detectors, plus the earlier 6 years
exposure of the original DAMA/NaI experiment with 100 kg of
detectors \cite{dm_dama10}, for a cumulated exposure of 1.17 t$\cdot$y. They
observe an annual modulation of the signal in the 2 to 6 keVee bin, with the
expected period (1 year) and phase (maximum around June 2), at 8.9 $\sigma$
level. If interpreted within the standard halo model described above, two
possible explanations have been proposed: a WIMP with $m_\chi \simeq 50$ GeV
and $\sigma_{\chi p} \simeq 7 \cdot 10^{-6}$ pb (central values) or at low
mass, in the 6 to 10 GeV range with $\sigma_{\chi p} \sim 10^{-3}$ pb; the
cross section could be somewhat lower if there is a significant channeling
effect \cite{dm_bertone}. 

Interpreting these observations as positive WIMP signal raises several issues
of internal consistency. First, the proposed WIMP solutions would induce a
sizeable fraction of nuclear recoils in the total measured rate in the 2 to 6
keVee bin. No pulse shape analysis has been reported by the authors to check
whether the unmodulated signal was detectable this way. Secondly, the residual
$e/\gamma$-induced background, inferred by subtracting the signal predicted by
the WIMP interpretation from the data, has an unexpected
shape \cite{dm_fairbairn}, starting near zero at threshold and quickly rising
to reach its maximum near 3 to 3.5 keVee; from general arguments one would
expect the background (e.g. due to electronic noise) to increase towards the
threshold. Finally, the amplitude of the annual modulation shows a somewhat
troublesome tendency to decrease with time.  The original DAMA
data, taken 1995 to 2001, gave an amplitude of the modulation of $20.0 \pm
3.2$ in units of $10^{-3}$ counts/(kg$\cdot$day$\cdot$keVee), in the 2-6 keVee
bin. During the first phase of DAMA/LIBRA, covering data taken between 2003
and 2007, this amplitude became $10.7 \pm 1.9$, and in the second phase of
DAMA/LIBRA, covering data taken between 2007 and 2009, it further decreased to
$8.5 \pm 2.2$. 
The ratio of amplitudes inferred from the DAMA/LIBRA phase 2 and
original DAMA data is $0.43 \pm 0.13$, differing from the expected
value of 1 by more than 4 standard deviations. (The results for the
DAMA/LIBRA phase 2 have been calculated by us using published
results for the earlier data alone \cite{dm_dama_previous} as well as
for the latest grand total \cite{dm_dama10}.) Similar conclusions can
be drawn from analyses of the 2-4 and 2-5 keVee bins.

Concerning compatibility with other experiments (see below), the high
mass solution is clearly excluded by several null observations (CDMS,
EDELWEISS, XENON), while possibly a small parameter space remains
available for the low mass solution (according to \cite{dm_fairbairn}
this possibility is excluded if the energy spectrum measured by
DAMA/LIBRA is taken into account). It should be noted that these
comparisons have to make assumptions about the WIMP velocity
distribution (see above), but varying this within reasonable limits
does not resolve the tension \cite{dm_fairbairn}. Moreover, one usually
assumes that the WIMP scatters elastically, and that the
spin-independent cross section for scattering off protons and neutrons
is roughly the same. These assumptions are satisfied by all models we
know that are either relatively simple (i.e. do not introduce many new
particles) or have independent motivation (e.g. attempting to solve
the hierarchy problem). As noted earlier, recently models have been
constructed where these assumptions do not hold, but at least some of
these are no longer able to make the WIMP interpretation of the
DAMA(/LIBRA) observations compatible with all null results from other
experiments. Finally, appealing to spin-dependent interactions does
not help, either \cite{dm_copi}, in view of null results from direct
searches as well as limits on neutrino fluxes from the Sun (see the
subsection on indirect WIMP detection below).

No other annual modulation analysis with comparable sensitivity has
been reported by any experiment. ANAIS \cite{dm_portal}, a 100 kg
NaI(Tl) project planned to be run at the Canfranc lab, is in the phase
of crystal selection and purification. DM-ice is a new project with
the aim of checking the DAMA/LIBRA modulation signal in the southern
hemisphere. It will consist of 250 kg of NaI(Tl) installed in the
heart of the IceCube array. The counting rate of crystals from the
previous NAIAD array recently measured in situ is currently dominated
by internal radioactivity.

KIMS \cite{dm_kims_web}, an experiment operating 12 crystals of CsI(Tl) with a
total mass of 104.4 kg in the Yang Yang laboratory in Korea, has accumulated
several years of continuous operation. They should soon be able to set an
upper limit on annual modulation amplitude lower than DAMA value if no annual
modulation is present, or confirm the DAMA value at 3 $\sigma$.

At mK temperature, the simultaneous measurement of the phonon and ionization
signals in semiconductor detectors permits event by event discrimination
between nuclear and electronic recoils down to 5 to 10 keV recoil energy. This
feature is being used by the CDMS \cite{dm_portal} and
EDELWEISS \cite{dm_portal} collaborations. Surface interactions, exhibiting
incomplete charge collection, are an important residual background, which is
treated by two different techniques: CDMS uses the timing information of the
phonon pulse, while EDELWEISS uses the ionization pulses in an interleaved
electrodes scheme. New limits on the spin-independent coupling of WIMPs were
obtained by CDMS, after operating 19 Germanium cryogenic detectors at the
Soudan mine during new runs involving a total exposure of around 612
kg$\cdot$d (around 300 kg$\cdot$d fiducial) \cite{dm_cdms2011}. Two events
were found in the pre-defined signal region, while 0.9 background event were
expected. Given these figures, no observation of a signal is claimed. While
this data set alone provided a worse limit than the previous runs, the
combined data sets provide an improved upper limit on the spin-independent
cross section for the scattering of a 70 GeV/c$^2$ WIMP on a nucleon of
3.8$\times 10^{-8}$ pb, at 90\% CL. The {\it ``WIMP safe''} minimal mass (see
the discussion at the end of sec.~1.2.4) of this analysis is about 12 GeV.

An independent analysis of data at low energy (i.e. above 2 keV recoil
energy) has also been performed by CDMS \cite{dm_cdms2011_le}. From the
knowledge of the quenching factor of Germanium recoils down to 2 keV
recoil energy, the energy spectrum is reconstructed using only the
measured phonon energy. The obtained spectrum, once corrected for
quenching, has a shape somewhat similar to that reported by CoGeNT,
but with a lower amplitude (especially for one of the detector
modules, which was used to set the limit) so that CDMS concludes that
their data are inconsistent with the original WIMP interpretation of
the CoGeNT data (note that both detectors use the same target
material, so this comparison really is model-independent), as well as
with the standard WIMP interpretation of the DAMA data. New detectors
with interleaved electrode schemes are being built.

EDELWEISS has operated ten 400 g Germanium detectors equipped with different
thermal sensors and an interdigitised charge collection electrode scheme,
during one year at the Laboratoire Souterrain de Modane \cite{dm_edw2011}. A
total of 5 events were observed in the signal region for a fiducial exposure
of 384 kg$\cdot$d, while 3 events were expected from backgrounds. No WIMP
signal was claimed. A similar sensitivity to CDMS is obtained at high mass,
while the high 20 keV analysis threshold induces a somewhat poorer limit at
masses lower than 50 GeV. New larger detectors with a complete coverage of the
crystal with annular electrodes, and better rejection of non-recoil events
are being built.

Given their similar sensitivities, the two collaborations combined their data
sets. Using a simple combination method, a gain of 1.6 relative to the best
limit has been obtained at WIMP masses larger than 700 GeV, 
and an improved limit of 3.3$\times 10^{-8}$ pb
for a 90 GeV WIMP mass \cite{dm_cdmsedw2011}.

The cryogenic experiment CRESST \cite{dm_portal} uses the scintillation
of CaWO$_{\rm 4}$ as second variable for background discrimination.
CRESST has recently submitted for publication \cite{dm_cresst2011} the
result of the analysis of 730 kg$\cdot$d exposure performed with 8
detectors. The observation of 67 events in the signal region does not
match the about 40 expected background events, originating from
e/$\gamma$ leakage, neutron recoils, as well as leakage from $\alpha$
and Pb recoils. The event excess is said to be compatible with
WIMPs. A likelihood method provides two solutions, respectively for 12
and 25 GeV masses, stating also that the background hypothesis alone
is more than 4 sigma away from the observed data. However, some other
potential sources of background are insufficiently addressed, like
``no-light'' events, a category of events which previously plagued the
sensitivity of this experiment.
  
Other inorganic scintillators are also being explored, e.g. by the
ROSEBUD collaboration \cite{dm_portal}.

The experimental programs of CDMS II, EDELWEISS II and CRESST II aim at an
increase of sensitivity by a factor of 10, by operating around 40 kg of
detectors. The next stage SuperCDMS and EURECA-I (a combination of EDELWEISS
and CRESST) projects will involve typically 150 kg of detectors. Then GEODM
and EURECA-2 will turn to 1 t goals.

Noble gas detectors for dark matter detection are now being developed rapidly
by several groups \cite{dm_bertone}. Dual (liquid and gas) phase detectors
allow to measure both the primary scintillation and the ionization electrons
drifted through the liquid and amplified in the gas, which is used for
background rejection.

The XENON collaboration \cite{dm_portal} has successfully operated the 161 kg
XENON100 setup at Gran Sasso laboratory during a 100 day data taking
period. Within a fiducial mass of 48 kg, 3 events were observed in the signal
region, while 1.8 were expected, out of which 1.2 originate from a sizeable
contamination of Krypton 85 in the liquid \cite{dm_xenon100}. This allowed to
set the best limits at all masses on spin-independent interactions of WIMPs,
with a minimum of cross section at 7.0$\times 10^{-9}$ pb for a mass of 50
GeV. However, the reliability of limits set at masses lower than 10 GeV,
especially wrt the relative light efficiency factor, have been discussed in
the community. Moreover, as underlined near the end of section 1.2.4, the
limits at low mass can be set {\it only} thanks to the poor energy resolution
at threshold --8.4 keV-- due to the low photoelectron yield of 0.5
pe/keV. With infinite energy resolution, a Xe detector {\it with the
  same threshold of 8.4 keV} is not sensitive to a WIMP mass of 7 GeV. 
Folding in the XENON100
resolution, the expected fraction of a 7 GeV WIMP signal above 8.4 keV is
around 0.05~\% (in strong contrast with the 10 \% to which CoGeNT is
sensitive). If one follows the recommendation made above, the {\it ``WIMP
safe''} minimal mass for XENON100 is around 12 GeV.

A reanalysis of part of the XENON10 data \cite{dm_xenon10le}, using the
ionization signal only, with an ionization yield of around 3.5
electron/keV at a threshold of 1.4 keV, sets a more convincing limit
in the 7 GeV range, about one order of magnitude below the original
CoGeNT claim (see above). The {\it "WIMP safe"} minimal mass for this
XENON10 analysis is around 5 GeV. The XENON10 limit for spin dependent
WIMPs with pure neutron couplings is still the best published limit at
all masses \cite{dm_xenonsd} (but likely to be soon superseded by an
analysis of XENON100 data). XENON1t, the successor of XENON100
planned to be run at Gran Sasso lab, is in its preparation phase.  One
should note that, presumably, the planned increase of distance 
between planes of PMT's will lead to a lower photoelectron yield 
for scintillation light than at XENON100. This was the case when 
going from XENON10 (around 1 pe/keV) to XENON100 (around 
0.5 pe/keV). For comparison, a 0.25 pe yield per keV 
would correspond to a {\it ``WIMP safe''}  mass of order of 20 GeV.	 

A new liquid Xenon based project, PANDA-X, with pancake geometry, planned to
be housed in the new Jinping lab, will perform a dedicated low mass WIMP
search.

ZEPLIN III \cite{dm_portal}, using a similar principle and with an
active mass of 12 kg of Xenon, operated in the Boulby laboratory, has
been upgraded for a lower background, has acquired new data, and is
now stopped. XMASS \cite{dm_portal} in Japan is close to operate a
single-phase 800 kg detector (100 kg fiducial mass) installed in a
large pure water shield at the SuperKamiokande site. With no pulse
shape analysis, the expected performance relies heavily on the
self-shielding effect to lower the background \cite{dm_bertone}.

The LUX detector \cite{dm_bertone}, a 300 kg double phase Xenon detector,
planned to be operated in the new SURF (previous Sanford) laboratory in US, is
in the commissioning phase, in a water shield at surface, before transport
underground to the 4850 level.

The WARP collaboration \cite{dm_portal} is currently installing a 100 l
Argon detector at the Gran Sasso laboratory. Thanks to a
double-background rejection method based on the asymmetry between
scintillating and ionizing pulses and extremely efficient pulse shape
discrimination of scintillating pulses, it looks possible to achieve
very high background rejection, even in the presence of the
radioactive isotope ${}^{39}$Ar. The ArDM project \cite{dm_portal} is
using a similar technique with a much larger (1,100 kg) mass. It
should be installed soon and take data at the newly opened Canfranc
laboratory. MiniCLEAN and DEAP-3600 \cite{dm_portal}, both measuring
only scintillation signals in spherical geometries in single phase
mode, are being assembled at SNOLab and will operate respectively 500
kg of Ar/Ne and 3600 kg of Ar \cite{dm_bertone}. DARK SIDE \cite{dm_portal}, 
is another  Argon based, double phase project,  involving in a first step 
about 50 kg of ${}^{39}$Ar depleted Argon, to be installed in Gran Sasso lab.

The low pressure Time Projection Chamber technique is the only convincing way
to measure the direction of nuclear recoils and prove the galactic origin of a
possible signal \cite{dm_bertone}. The DRIFT collaboration \cite{dm_portal} has
operated a 1 m$^3$ volume detector in the UK Boulby mine. Though the
background due to internal radon contamination was lowered, no new competitive
limit has yet been set. The MIMAC collaboration \cite{dm_portal} is
investigating a sub-keV energy threshold TPC detector. Additional sensitive
measurements of Fluor nuclei quenching factor and recoil imaging have been
performed recently by this group down to few keV. A 2.5 l 1000 channel
prototype is going to be operated soon in the Fr\'ejus laboratory. Other
groups developing similar techniques, though with lower sensitivity, are DMTPC
in the US and NewAge in Japan.

\begin{figure}[h!]
\centering\includegraphics[width=1.\textwidth]{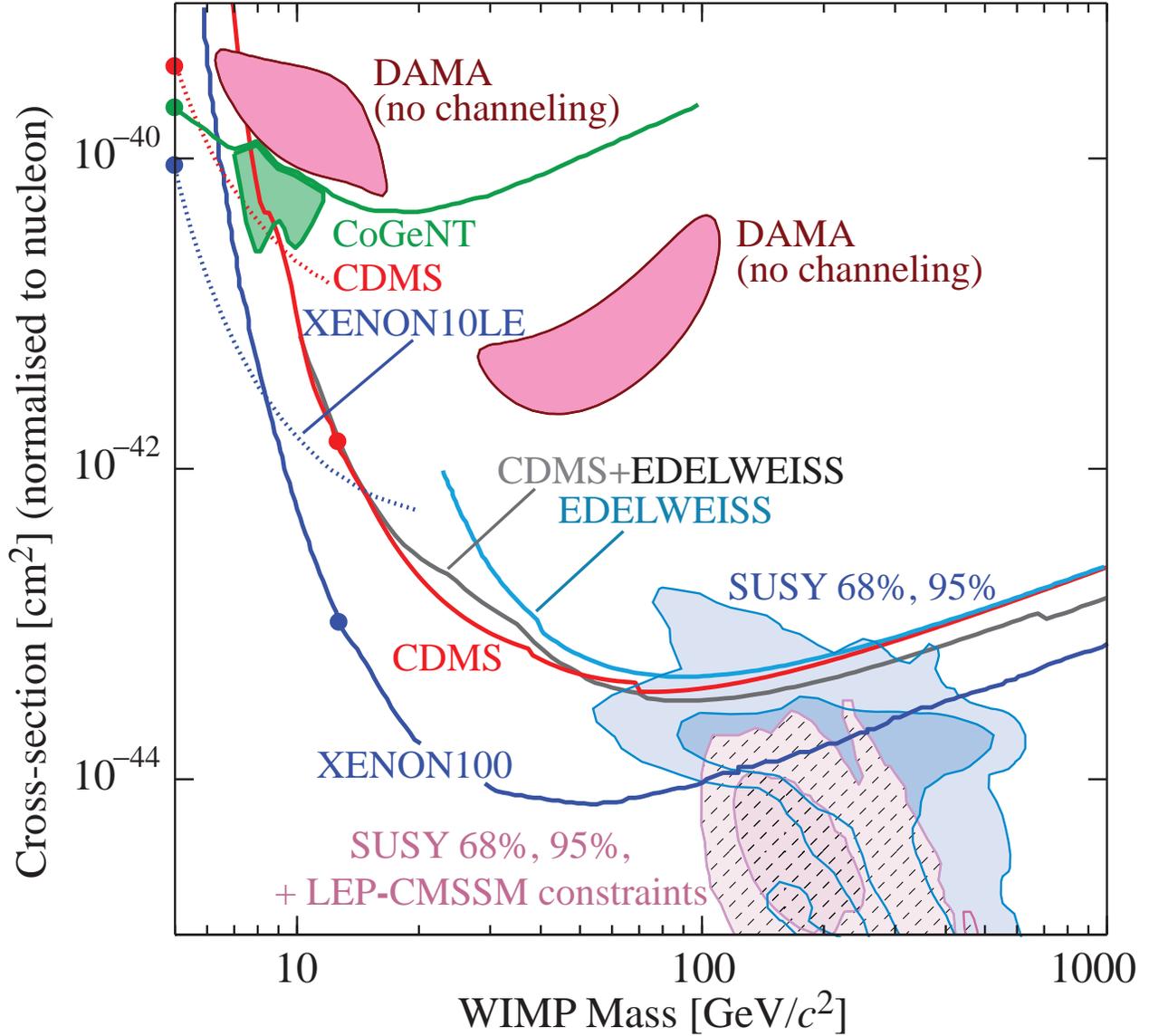}

\caption{Cross sections (normalized to nucleon
assuming $A^{2}$ dependence, see section 1.2.4) for spin independent
coupling versus mass diagrams. Please refer to the text to get the
references of corresponding publications of experimental results. The
big dots on some curves show the {\it ``WIMP safe"} minimal mass for
the corresponding experimental result (see details in text).  DAMA
candidates region (no channeling) are from \cite{dm_DAMA_savage},
shaded 68 and 95 \% regions are SUSY predictions by
\cite{dm_trotta2008}, together with recent constraints (light gray 68
and 95 \% contours) placed by LHC experiments, both on the CMSSM
\cite{dm_lhc}. Here equal cross sections for scattering off protons
and neutrons have been assumed.}
\end{figure}

The following more unconventional detectors use $\rm ^{19}$F nuclei to
set limits on the spin dependent coupling of WIMPs, with less than kg
mass detectors. The bubble chamber like detector,
COUPP \cite{dm_portal}, run at Fermilab, has provided a new
limit \cite{dm_coupp2} for spin dependent proton coupling WIMPs for
masses above 20 GeV, superseding an earlier KIMS
result. PICASSO \cite{dm_portal}, a superheated droplet detector run at
SNOLAB, obtained a better limit below 20 GeV on the same type of
WIMPs \cite{dm_picasso}. Finally, SIMPLE \cite{dm_portal}, a similar
experiment run at Laboratoire Souterrain de Rustrel, submitted results
for publication that claim to provide the currently best limit on the
spin-dependent WIMP-proton cross section for all WIMP
masses \cite{dm_simple}.

Figures 1 and 2 illustrate the above results on limits on and positive
claims of cross sections, normalized to nucleon, 
for spin independent and spin dependent
couplings, respectively, as functions of WIMP mass, where only the two
currently best limits are presented. Also shown are constraints from
indirect observations (see the next section) and typical regions of
SUSY models, before and after recent LHC results. These figures have
been made with the {\tt dmtools} web page \cite{dm_dmtools}.

\begin{figure}[t!]
\centering\includegraphics[width=1.\textwidth]{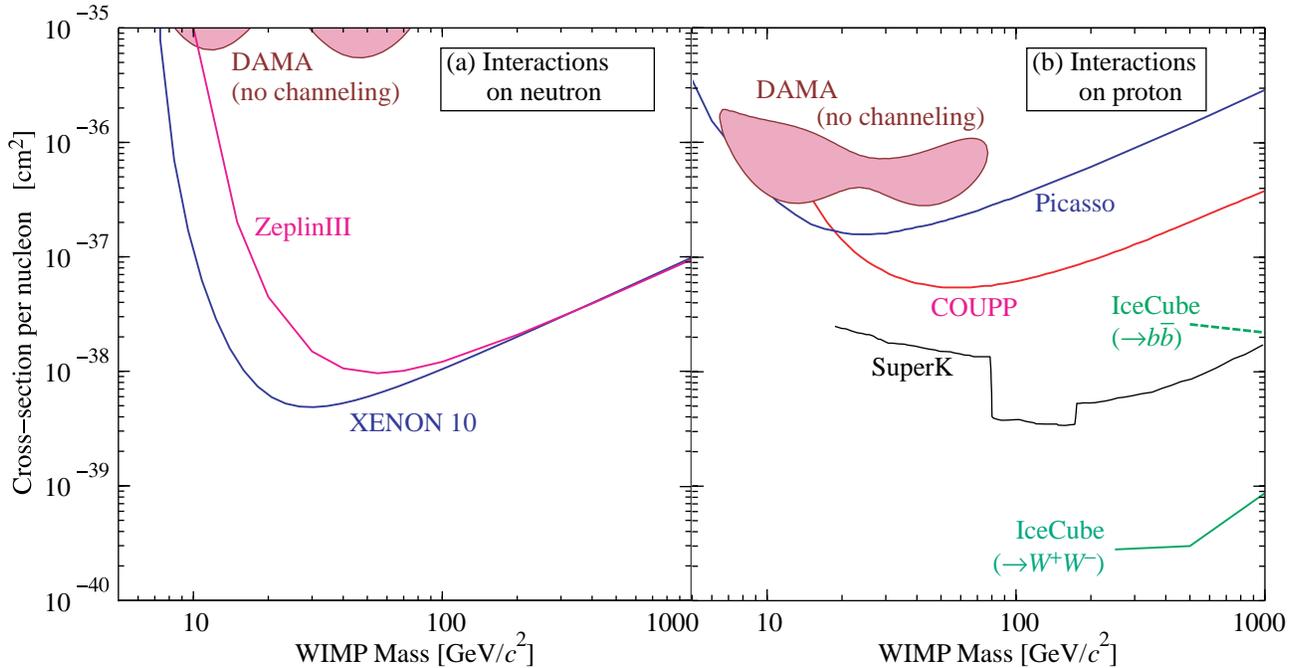}

\caption{Cross sections for spin dependent coupling
versus mass diagrams. Please refer to the text to get the references
of corresponding publications of experimental results.  DAMA
candidates region (no channeling) are from \cite{dm_DAMA_savage} Left:
interactions on neutron. Right: interactions on proton.}
\end{figure}

Sensitivities down to $\sigma_{\chi p}$ of $10^{-10}$ pb, as needed to probe
large regions of MSSM parameter space \cite{dm_mssm}, will be reached with
detectors of typical masses of 1 ton, assuming nearly perfect background
discrimination capabilities. Note that the expected WIMP rate is then 5
evts/ton/year for Ge. The ultimate neutron background will only be identified
by its multiple interactions in a finely segmented or
multiple-interaction-sensitive detector, and/or by operating detectors
containing different target materials within the same set-up. Larger mass
projects are envisaged by the DARWIN European consortium and the MAX project
in the US (liquid Xe and Ar multiton project) \cite{dm_portal}.

\subsection{Status and prospects of indirect WIMP searches}

WIMPs can annihilate and their annihilation products can be detected;
these include neutrinos, gamma rays, positrons, antiprotons, and
antinuclei \cite{dm_bertone}. These methods are complementary to direct
detection and might be able to explore higher masses and different
coupling scenarios.  ``Smoking gun'' signals for indirect detection
are GeV neutrinos coming from the center of the Sun or Earth, and
monoenergetic photons from WIMP annihilation in space.

WIMPs can be slowed down, captured, and trapped in celestial objects like the
Earth or the Sun, thus enhancing their density and their probability of
annihilation. This is a source of muon neutrinos which can interact in the
Earth. Upward going muons can then be detected in large neutrino telescopes
such as MACRO, BAKSAN, SuperKamiokande, Baikal, AMANDA, ANTARES, NESTOR, and
the large sensitive area IceCube \cite{dm_bertone}. The best upper limit for
relatively soft muons, of $\simeq$ 1000 muons/km$^2$/year for muons with
energy above $\sim 2$ GeV \cite{dm_sk}, comes from
SuperKamiokande \cite{dm_portal} using through-going muons. For more energetic
muons a slightly more stringent limit has been set by IceCube22 (using 22
strings), e.g. excluding a flux above 610 muons/km$^2$/year from the Sun for a
WIMP model with average muon energy of 150 GeV \cite{dm_ice22}. In the
framework of the MSSM and with standard halo velocity profiles, only the
limits from the Sun, which mostly probe spin-dependent couplings, are
competitive with direct WIMP search limits. IceCube80 \cite{dm_portal} will
increase this sensitivity by a factor $\simeq$ 5 at masses higher than 200 GeV
while IceCube Deep Core will allow to reach masses down to 50
GeV \cite{dm_bertone}. 

WIMP annihilation in the halo can give a continuous spectrum of gamma
rays and (at one-loop level) also monoenergetic photon contributions
from the $\gamma \gamma$ and $\gamma Z$ channels.  These channels also
allow to search for WIMPs for which direct detection experiments have
little sensitivity, {\it e.g.}, almost pure higgsinos.  However, the
size of this signal depends very strongly on the halo model, but is
expected to be most prominent near the galactic center. The central
region of our galaxy hosts a strong TeV point source
discovered \cite{dm_hess_cen} by the H.E.S.S. Cherenkov
telescope \cite{dm_hess_web}. Moreover, FERMI/LAT \cite{dm_portal} data
revealed a new extended source of GeV photons near the galactic center
above and below the galactic plane \cite{dm_bubbles}. Both of these
sources are most likely of astrophysical origin. The presence of these
unexpected backgrounds makes it more difficult to discover WIMPs in
this channel, and no convincing signal has been claimed. FERMI/LAT
observations of the galactic halo are in agreement with predictions
based on purely astrophysical sources (in contrast to a re-analysis of
earlier EGRET data \cite{dm_egret2}), and rule out many WIMP models
that were constructed to explain the PAMELA and FERMI/LAT excesses in
the $e^\pm$ channel \cite{dm_lat_halo}. Similarly, Cherenkov telescope
and FERMI/LAT observations of nearby dwarf galaxies, globular
clusters, and clusters of galaxies only yielded upper limits on photon
fluxes from WIMP annihilation. While limits from individual
observations are still above the predictions of most WIMP models, a
very recent combination \cite{dm_dwarf_comb} of limits from dwarf
galaxies excludes WIMPs annihilating hadronically with the standard
cross section needed for thermal relics, if the WIMP mass is below 25 GeV;
assumptions are annihilation from an $S-$wave initial state, and a
dark matter density distribution scaling like the inverse of the
distance from the center of the dwarf galaxy at small radii.

Antiparticles arise as additional WIMP annihilation products in the
halo. To date the best measurement of the antiproton flux comes from
the PAMELA satellite \cite{dm_portal}, and covers kinetic energies
between 60 MeV and 180 GeV \cite{dm_pamelaantip}. The result is in good
agreement with secondary production and propagation models. These data
exclude WIMP models that attempt to explain the $e^\pm$ excesses via
annihilation into $W^\pm$ or $Z^0$ boson pairs; however, largely due
to systematic uncertainties they do not significantly constrain
conventional WIMP models.

The best measurements of the positron (and electron) flux at (tens of)
GeV energies again comes from PAMELA \cite{dm_pamelaelec}, showing a
rather marked rise of the positron fraction between 10 and 100
GeV. The observed spectrum falls within the one order of magnitude
span (largely due to differences in the propagation model used) of
positron fraction values predicted by secondary production
models \cite{dm_modelselec}.  Measurements of the total
electron+positrons energy spectrum by ATIC \cite{dm_atic},
FERMI/LAT \cite{dm_fermielec} and H.E.S.S. \cite{dm_hesselec} between
100 and 1000 GeV also exceed the predicted purely secondary spectrum,
but with very large dispersion of the magnitude of these
excesses. While it has been recognized that astrophysical sources may
account for all these features, many ad-hoc Dark Matter models have
been built to account for these excesses. As mentioned in section 1,
given the amount of jerking and twisting needed to build such models
not to contradict any observation, it seems very unlikely that Dark
Matter is at the origin of these excesses.

Last but not least, an antideuteron signal \cite{dm_bertone}, as potentially
observable by AMS2 or PAMELA, could constitute a signal for WIMP annihilation
in the halo.

An interesting comparison of respective sensitivities to MSSM
parameter space of future direct and various indirect searches has
been performed with the DARKSUSY tool \cite{dm_darksusy}. A web-based
up-to-date collection of results from direct WIMP searches,
theoretical predictions, and sensitivities of future experiments can
be found in \cite{dm_dmtools}. Also, the web page \cite{dm_iliasdm}
allows to make predictions for WIMP signals in various experiments,
within a variety of SUSY models and to extract limits from simply
parametrised data. Integrated analysis of all data from direct and
indirect WIMP detection, and also from LHC experiments should converge
to a comprehensive approach, required to fully unravel the mysteries
of dark matter.

\subsection*{Acknowledgments}
We thanks Elena Aprile, Keith Olive and Alessandro Strumia for useful
comments, and Don Groom for help with the graphics. Special thanks to
the members of the DMtools team, who made Figs.~1 and 2 possible.

\end{document}